## И. З. ШКУРЧЕНКО

### ПРИМЕРЫ ПРИМЕНЕНИЯ МЕХАНИКИ МЕХАНИКИ БЕЗЫНЕРТНОЙ МАССЫ

Эти примеры отсутствуют в тексте «Механики жидкости и газа, или механики безынертной массы», о чём сказано в «Предисловии редактора» (Часть I данной монографии). Редактор поместил недостающий текст отдельно, так как автор доработал некоторые теоретические и практические положения, включая большинство примеров, в своих последующих исследованиях. Поэтому теперь «Примеры» лучше читать после прочтения всех текстов по теме механики безынертной массы, находящихся в этом архиве. Редактор больше не имеет других текстов автора по данной теме. «Примеры» адресованы специалистам в области теоретической и практической гидромеханики и смежных с ней наук.

### I. Z. SHKURCHENKO

### EXAMPLES OF APPLICATION OF MECHANICS OF THE INERTLESS MASS

These examples are absent in the text of "Mechanics of liquids and gas, or mechanics of the inertless mass", as it was said in "Foreword of the editor" (Part I of this monograph). The editor places the missing text separately since some theoretical and practical positions, including most examples, were developed by the author himself in two his next investigations. Therefore it will be more useful to read "The Examples" after all the monographs on mechanics of the inertless mass, which are in the same archive now. The editor has no other texts of the author on this subject. "The Examples" are addressed to specialists in the field of theoretical and practical hydrodynamics and adjacent sciences.

# СОДЕРЖАНИЕ:

**Стр.**

ПРИМЕР №1[1]

Требуется определить состояние жидкостей и газов в различных резервуарах и объёмах.

Здесь имеется в виду такое состояние жидкостей и газов, когда практически говоря, что они находятся под давлением. С точки зрения механики жидкости и газа, это состояние определяется как состояние покоя жидкостей и газов в скалярном силовом поле. Это значит, что жидкости и газы заполняют полностью весь объём резервуара и что давления в любых точках этого объёма одинаковы и равны по величине.

Статические силы давления в этих резервуарах определяются в зависимости от состояния массы жидкости или газа. Непосредственно силы давления определяются в соответствии с законом Паскаля, то есть, например, жидкости в резервуар нагнетаются под определённым давлением. Для газов силы давления можно определить, помимо закона Паскаля, ещё по уравнению состояния Клапейрона:

$$Pv = RT.$$

Потенциальная энергия этих объёмов определяется как произведение сил давления на объём. Если отнести энергию к единице объёма, то количество единиц сил давления будет равно количеству единиц потенциальной энергии. Это значит, что стрелка манометра остановится на одном и том же делении шкалы как при замере сил давления, так и при замере потенциальной энергии. При одинаковой величине потенциальной энергии жидкости и газа в случае преобразования энергии жидкости в работу мы не получим почти никакой работы. Для газа мы получим работу, величина которой определяется по уравнению энергии сжатия (VII.7). Как для установившегося вида движения, так и для состояния покоя, уравнение энергии механики жидкости и газа и показания манометра не учитывают энергию сжатия. Следовательно, характерная особенность уравнений энергии механики жидкости и газа.

В связи с тем, что мы рассматриваем состояние реальных жидкостей и газов, то мы должны учесть вязкость и термодинамическое состояние. Вязкость в данном случае не учитывается, так как отсутствуют динамические силы давления. Термодинамические условия притока или убыли тепловой энергии определяются с помощью уравнения энергии сжатия (VII.7). Тогда действительная величина потенциальной энергии состояния покоя жидкостей и газов $U_д$ запишется таким равенством:

$$U_д = U \pm \int_1^{V_2} P dV \,, \qquad (1)$$

где $U_д$ – действительная энергия объёма жидкости или газа, $U$ – первоначальная энергия объёма жидкости или газа, $\int_1^{V_2} P dV$ – потеря или прибыль энергии, связанная с термодинамическими условиями.

При вычислении энергии сжатия для записи термодинамических условий пользуются зависимостями термодинамических процессов. Отметим, что для жидкостей в таких случаях не вычисляют энергию сжатия, а пользуются коэффициентом линейного расширения, который по сути своей не отражает истинной количественной зависимости. Поэтому он затрудняет и усложняет подобные практические расчёты. Следовательно, для жидкостей должны быть получены зависимости типа зависимостей термодинамических процессов газа, чтобы для жидкостей можно было пользоваться уравнением энергии сжатия (VII.7).

Мы записали состояние жидкостей и газов для определённых замкнутых объёмов, которые определяются статическими силами давления и потенциальной энергией. Резервуар как конкретное условие при практическом использовании жидкостей и газов проявляет себя как замкнутый объём известной величины и как средство внешнего воздействия на жидкости и газы.

Отметим, что уравнение состояния газов Клапейрона с точки зрения механики жидкости и газа является уравнением энергии единицы веса газа, так как удельный объём есть объём единицы веса газа, а произведение этого объёма на давление выражает количественно энергию.

ПРИМЕР №2

Требуется определить взаимодействие жидкостей и газов, находящихся в состоянии покоя под действием векторного силового поля, с поверхностью твёрдого тела.

За векторное силовое поле в данном примере принимаем поле земного тяготения, под действием которого находятся жидкости. В таком состоянии они находятся либо в искусственных водоемах, либо в естественных. Твёрдые предметы также могут находиться либо в толще воды, либо плавать на ее поверхности. Во всех случаях их механическое взаимодействие с жидкостью происходит непосредственно через поверхность твёрдых тел, которая непосредственно соприкасается с жидкостью. Эта поверхность является не только поверхностью взаимодействия, но и границей раздела двух механик: механики твёрдого тела и механики жидкости и газа.

Механическое взаимодействие будет заключаться в том, что все твёрдые тела в данном примере будут находится под силовым воздействием жидкости на поверхности соприкосновения этих тел с жидкостью.

В данном примере нам необходимо определить зависимостями это силовое взаимодействие. Покажем на рисунке 28 часть воображаемого бассейна. На этом рисунке показано, что поверхностью соприкосновения жидкости и твёрдого тела служит днище и стенка до высоты h воображаемого бассейна. На нем также показаны две системы отсчета. Для механики жидкости и газа ею служит поверхность жидкости, относительно которой определяется высота h. Для механики твёрдого тела ею является декартова система координат. Мы разместили ее так, чтобы одна из осей совпадала с направлением высоты, а другая бы лежала в плоскости поверхности жидкости.

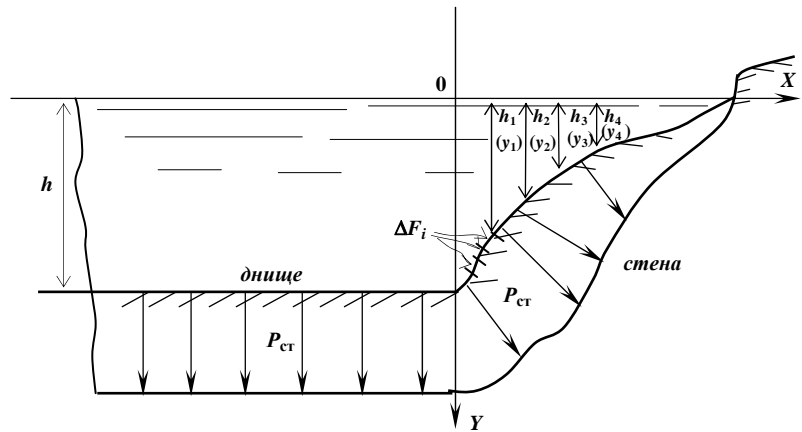

Рис. 28

В соответствии с положением о точке среды силы давления жидкости будут направлены в любой точке поверхности соприкосновения перпендикулярно к этой поверхности. Так как силы давления являются скалярными величинами, то направление их действия определяется поверхностью или плоскостью. Величина сил давления определяется по уравнению (II.15) в зависимости от высоты h. В свою очередь, действие сил давления жидкости уравновешивается на поверхности соприкосновения силовыми напряжениями, которые возникают в твёрдых телах. По отношению к твёрдому телу давление жидкости определяется как распределенные по поверхности силы. Распределение и сложение их в твёрдом теле определяется положениями механики твёрдого тела, которая рассматривает силы как векторные величины.

Дальше мы определяем распределение сил давления в бассейне и составляем условие равновесия. Силы, действующие со стороны твёрдого тела, будем определять как силу R, которая является функцией напряжений δ, то есть Rf(δ).

Теперь рассмотрим действие сил давления жидкости на днище и стенки бассейна. Принимаем, что днище расположено параллельно нулевой поверхности жидкости, тогда величина сил, действующих на днище, будет равна произведению площади днища F, умноженному на величину сил давления, которые определяются по уравнению (II.15). Эта сила будет уравновешиваться силой, действующей со стороны твёрдого тела на поверхность днища (т.е. поверхность твёрдого тела). Запишем это условие:

$$Rf(\delta) = F\rho h w^2. \qquad (2)$$

Вспомним, что в поле земного тяготения квадрат скорости $w^2$ в поле земного тяготения равен по абсолютной величине ускорению g в поле земного тяготения. Уравнением (2) мы записали условие равновесия для днища бассейна.

Принимаем, что профиль стенки бассейна по отношению к изменению высоты h может быть любым. Поэтому для определения сил, действующих на стенку, воспользуемся приближенным методом. Для чего разобьем площадь стенки на ряд элементарных площадей, где можно приближенно считать, что высота $h_j$ для всей элементарной площадки одинакова. Тогда для каждой элементарной площадки можно записать условие равновесия в таком виде:

$$\Delta R_i f(\delta) = \Delta F_i \rho w^2 h_j. \qquad (3)$$

Для всей поверхности бассейна условие равновесия запишется как сумма сил, действующих на всех элементарных площадках:

$$\sum_{i=1}^{n} R_i f(\delta) = \sum_{\substack{i=1 \\ j=1}}^{n} \Delta F_i \rho w^2 h_j. \qquad (4)$$

Эта сумма определит нам условие равновесия по отношению к неинерциальной системе координат, то есть по отношению к действующим силам давления жидкости. По отношению к твёрдому телу и инерциальной системе координат сумму (4) надо записать с учетом направления действия сил давления и их равнодействующей, то есть как сумму проекций на оси инерциальной системы координат. В общем, действия над ними производятся в соответствии с положениями механики твёрдого тела. Таким образом осуществляется переход и связь между положениями механики твёрдого тела и механики жидкости и газа.

Поверхность соприкосновения жидкости с твёрдым телом как одно из условий практического использования жидкостей и газов определяет в данном случае границу между положениями механики твёрдого тела и механики жидкости и газа.

ПРИМЕР №3

Требуется определить движение установившегося потока жидкости или газа в трубопроводах.

Для установившегося вида движения жидкостей и газов получены все необходимые зависимости. Согласно этим зависимостям, площадь сечения потока по участкам движения может быть различной. Они же определяют минимально возможную площадь сечения потока. В то же время из практики мы знаем, что для сохранения установившегося потока необходимо осуществлять плавный переход потока между его участками с различной площадью сечения. При наличии в потоке резких, ступенчатых, переходов от одной площади сечения к другой происходит значительное падение энергии потока, связанное с его геометрическими характеристиками. То есть эти потери непосредственно связаны с геометрической формой трубопровода, т.к. в данном примере его геометрические характеристики определяют геометрические характеристики потока. Целью данного примера является определение необходимых зависимостей и положений для участков плавного перехода потока с одной площади сечения на другую.

Дальше принимаем, что сечением трубопровода является прямоугольник, что переход потока от одной площади сечения к другой может осуществляться двумя способами: непосредственным изменением стенок трубопровода и размещением в трубопроводе твёрдого тела определённой формы. На рисунке 29 изобразим поток с сечением, изменяемым в соответствии с двумя этими способами.

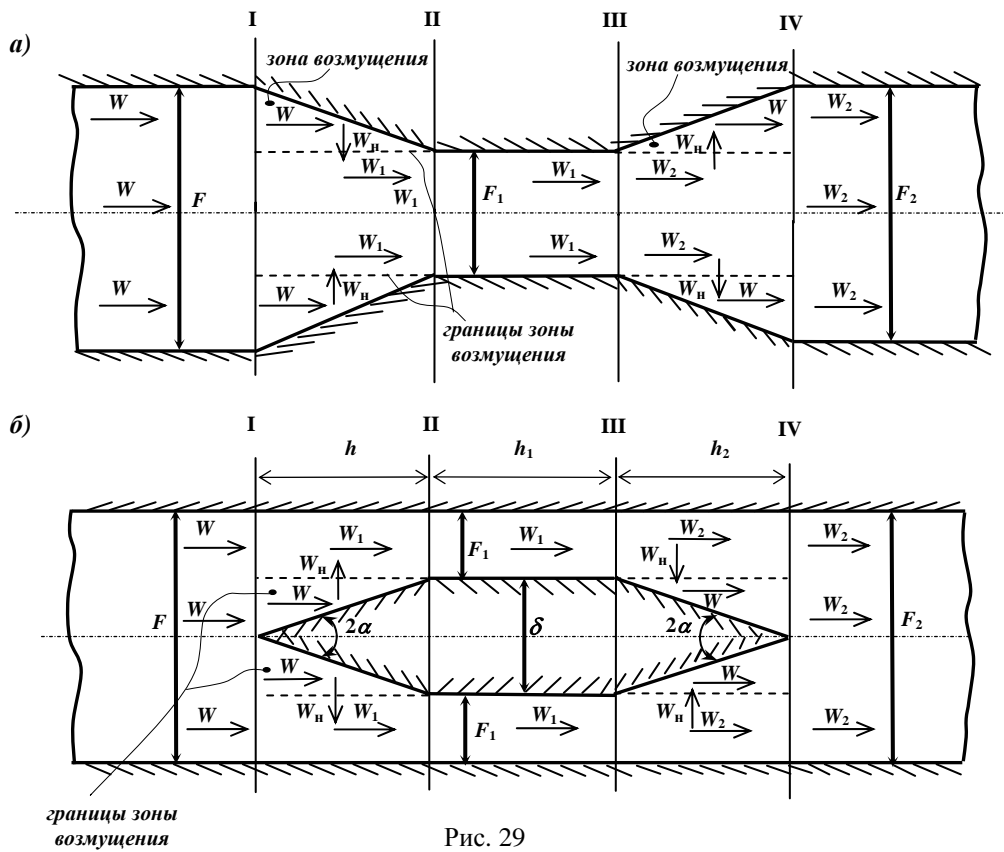

Рис. 29

Потоки, показанные на рисунках 29, *а* и 29, *б*, по расходу массы и по площадям сечения одинаковы. Установившийся поток требует симметрии для площадей своего сечения. По этой причине переход по сечению потока осуществляется за счет изменения формы его нижней и верхней сторон. Отклонение от требований симметрии приводит к определённым потерям энергии. Теперь переходим к непосредственному рассмотрению участков потока в порядке движения жидкости в потоке.

### Участок I -II

На этом участке происходит сужение потока. Такое изменение относится к пространственному изменению. Так как зависимости для установившегося движения не зависят от параметров пространства и времени, то для данного случая мы не можем воспользоваться ими. Поэтому нам придётся взять такие зависимости, которые связаны с пространственным изменением потока. Необходимые зависимости относятся к плоскому установившемуся виду движения.

Воспользуемся основным положением этого вида движения, которое утверждает, что пространственное движение потока происходит только в двух взаимно перпендикулярных направлениях. Одно направление нам известно – оно совпадает с направлением движения установившегося потока, то есть с направлением линии тока. Следовательно, второе направление движения будет иметь перпендикулярное к ней направление, то есть оно будет перпендикулярно оси потока. Теперь мы можем записать движение на участке потока I-II с помощью уравнений движения и сил. Дальнейшие рассуждения будем вести относительно рисунка 29, *б*, т.к. он образует одинаковую картину движения с рисунком 29, *а*, но более удобен для наглядного объяснения.

### Уравнения движения зоны возмущения

Границы зоны возмущения участка потока I-II определяются длиной участка $h$, толщиной центрального тела $\delta$, шириной потока *Ш*. Дополнительной характеристикой является угол $\alpha$ (рис. 29, *б*).

Объём участка I-II, который расположен вне зоны возмущения, назовём просто объёмом потока участка   I-II. Мы определили геометрические характеристики зоны возмущения.

Через плоскость I-I в зону возмущения будет поступать жидкость со скоростью $W$ и плотностью $\rho$. Плотность и скорость соответствуют величинам, которые определяются зависимостями предыдущего участка. Тогда уравнение движения для плоскости I-I можно будет записать в таком виде:

$$M_\text{п} = \rho W \cdot \delta\, Ш \qquad (1)$$

Уравнение (1) является уравнением движения зоны возмущения в плоскости I-I. С помощью уравнения (1) определяется поступательный расход массы в единицу времени $M_\text{п}$ в зоне возмущения. Этот расход вытеснит из зоны возмущения через ее границы в объём потока нормальный расход массы в единицу времени $M_\text{н}$. Нормальный и поступательный расходы массы зоны возмущения равны друг другу по абсолютной величине. В то же время нормальный расход массы будет отрицательным, т.к. он вытекает из зоны возмущения. Тогда уравнение для нормального расхода массы будет иметь вид:

$$M_\text{н} = -\rho W_\text{н} \cdot 2h Ш \qquad (2)$$

Уравнение (2) является уравнением движения нормального движения в зоне возмущения. Из этого уравнения мы можем определить нормальную скорость движения $W_\text{н}$ зоны возмущения.

Расход массы в единицу времени в объёме потока участка I-II будет равен расходу массы всего установившегося потока $M$, а площадь сечения равна площади сечения участка II-III. Следовательно, уравнение движения этого объёма будет одинаковым с уравнением движения участка III-III. Запишем его:

$$M = \rho W_1 F_1, \qquad (3)$$

где $W_1$ – скорость участка потока II-III, $F_1$ – площадь сечения потока на участке II-III.

Мы получили все необходимые уравнения движения участка потока I-II, которых три – (1), (2),(3).

### Уравнения сил участка потока I-II

Для плоскости I-I зоны возмущения уравнение сил будет иметь вид:

$$P_\text{пр.п} = P_\text{ст.п} + \rho W^2. \qquad (4)$$

Для нормального движения на границе зоны возмущения оно будет иметь вид:

$$P_\text{пр.н} = P_\text{ст.н} - \rho W_\text{н}^2. \qquad (5)$$

Для объёма потока участка I-II оно будет иметь вид:

$$P_{\text{пр}} = P_{\text{ст}1} + \rho W_1^2. \tag{6}$$

Уравнения (4), (5), (6) являются уравнениями сил участка потока I-II.

Величины сил давления в этих уравнениях вычисляются почленно. Сначала вычисляют динамические силы давления, так как плотность и скорость для каждого из них известны. Они были получены по уравнениям движения. Дальше, исходя из условия постоянства полной энергии потока для любой точки объёма участка I-II, определим величины статических сил давления в уравнениях (4), (5), (6). Затем мы получим принятые, или суммарные, силы давления $P_{\text{пр}}$ для каждого из них. Решив уравнения движения и сил участка I-II для конкретного установившегося потока жидкости, мы получим определённые количественные величины для него.

Теперь перейдем к исследованию, которое будет заключаться в том, что мы рассмотрим влияние изменения угла $\alpha$ на характеристики нормального движения зоны возмущения и всего движения в целом на участке потока I-II. Принимаем диапазон изменения угла $\alpha$ от 0° до 90°. При стремлении угла $\alpha$ к нулю длина $h$ участка потока I-II будет стремиться к бесконечности. Это значит, что в уравнении движения (2) для нормального расхода массы в зоне возмущения площадь нормального движения будет стремиться к бесконечности. В связи с тем, что расход массы является постоянной величиной, то нормальная скорость движения $W_{\text{н}}$ будет стремиться к нулю. Теперь полагаем, что угол $\alpha$ стремится к 90°. Тогда длина участка I-II $h$ и площадь нормального расхода в уравнении (2) будут стремиться к нулю. В связи с тем, что расход массы остаётся постоянной величиной, то нормальные скорости движения $W_{\text{н}}$ будут стремиться к бесконечности. При изменении угла $\alpha$ в диапазоне исследования другие характеристики потока участка I-II останутся без изменения, то есть уравнения (1), (3) и (4), (6) и их переменные останутся неизменными.

Данное исследование интересовало нас по такому вопросу: каковы могут быть максимальные и минимальные силы давления, действующие на площадь сечения центрального тела ($\delta \cdot Ш$) на участке потока I-II. Величина этих сил давления определяется уравнениями сил (4), (5) зоны возмущения этого участка.

При стремлении угла $\alpha$ к нулю нормальные динамические силы давления в уравнении (5) тоже стремятся к нулю. Это значит, что величина силы, действующей на площадь сечения центрального тела, будет определяться только уравнением (4), то есть поступательными динамическими и статическими силами давления. В этом случае мы будем иметь максимальные силы давления на площади сечения центрального тела.

При стремлении угла $\alpha$ к 90° нормальные динамические силы давления стремятся к бесконечности. Конечно, они не достигают этой величины, так как по условиям движения, которые были рассмотрены выше, динамические силы давления достигают своей максимальной величины, которая равна статическому давлению сил для полной энергии установившегося потока. Если угол $\alpha$ находится в близких пределах к максимальным динамическим силам давления, то в этом случае мы получим минимальные давления на площади сечения центрального тела.

Уменьшение сил давления здесь происходит за счет уменьшения статических сил давления в уравнении (5). Так как нормальные динамические силы давления действуют перпендикулярно направлению потока, то они не оказывают никакого прямого воздействия на центральное тело. Они действуют на него как бы косвенным путем. Их увеличение приводит к уменьшению статического давления в уравнении (5). В то же время величины этих статических сил давления не может быть больше величины статических сил давления, определённых по уравнению (4), что, в свою очередь, приводит к уменьшению сил, действующих на площадь сечения центрального тела, конечно, за счет уменьшения статических сил давления. Количественно это можно определить по распределению полной энергии потока между потенциальной и кинетической. Большие нормальные скорости, по сравнению с поступательными скоростями движения, приводят к уменьшению статических сил давления. Отсюда следует, что минимальные силы давления действуют на поперечное сечение центрального тела при максимальных нормальных скоростях движения в зоне возмущения участка потока I-II.

Следующее условие, которое нас тоже интересует, это обтекаемость, то есть движение потока на участке I-II без потерь полной энергии потока. Рассмотрим обтекаемость центрального тела в зависимости от его формы на участке потока I-II.

При стремлении угла $\alpha$ к нулю увеличивается длина профиля, но это не приводит к нарушениям в зависимостях сил, движения и энергии. Следовательно, никаких потерь полной энергии потока здесь не будет, то есть профиль центрального тела на участке потока I-II будет обтекаем.

При стремлении угла $\alpha$ к 90° профиль центрального тела на участке потока I-II стремится к плоскости II-II, а нормальные скорости – к бесконечности. Здесь мы наблюдаем нарушение распределения полной энергии потока между потенциальной и кинетической. В этом случае реальные, максимально возможные нормальные скорости поддерживаются частично за счет потерь полной энергии потока. Тогда подобный профиль относится к разряду плохо обтекаемых, так как он требует потерь полной энергии потока. Чтобы сделать профиль обтекаемым, выбирают угол $\alpha$ таким, чтобы он не нарушал условие распределения полной энергии потока между потенциальной и кинетической. Практика показывает, что обтекаемый профиль с окружностью вместо треугольника соответствует профилю с максимально возможным углом $\alpha$.

Мы рассмотрели условие обтекаемости профиля. Будем считать, что на этом задачу по участку потока I-II мы выполнили.

Участок потока II-III есть обыкновенный участок установившегося потока. Поэтому для него пригодны все зависимости установившегося вида движения. В связи с этим мы не будем его рассматривать.

***Рассмотрим движение на участке потока III - IV.***

Сначала сопоставим картину движения на участках потока I-II и III-IV. Принципиальное различие здесь будет заключаться в том, что скорости движения в зоне возмущения этих участков будут иметь противоположные направления. Характер изменения движения на участке III-IV происходит относительно последующего участка потока. Поэтому непосредственно в объёме потока участка III-IV скорость будет равна скорости на последующем участке $W_2$. Величина расхода массы в единицу времени зоны возмущения определяется в плоскости IV-IV для площади сечения зоны возмущения по уравнению движения. Запишем его:

$$M_п = -\rho W \cdot \delta Ш \qquad (7)$$

Для нормального движения на границе зоны возмущения уравнение движения будет иметь такой вид:

$$M_н = \rho \, W_н \cdot h_2 \ Ш. \qquad (8)$$

Расходы масс в уравнениях (7) и (8) равны друг другу, то есть $M_п = M_н$. Скорость движения в объёме потока участка III-IV определяется по уравнению движения следующего участка, поэтому

$$M = \rho W_2 F_2. \qquad (9)$$

Мы получили все необходимые уравнения движения для участка потока III-IV.

Запишем теперь уравнения сил для этого участка.

Уравнение сил поступательного движения зоны возмущения участка III-IV будет иметь вид:

$$P_{пр.п} = P_{ст.п} - \rho W^2. \qquad (10)$$

Уравнение сил нормального движения зоны возмущения участка III-IV будет иметь вид:

$$P_{пр.н} = P_{ст.н} + \rho W_н^2. \qquad (11)$$

Уравнение сил объёма потока участка III-IV будет иметь вид:

$$P_{пр} = P_{ст2} + \rho W_2^2. \qquad (12)$$

Мы получили уравнения сил для участка потока III-IV. Они тоже вычисляются почленно, как и для участка I-II.

При изменении угла $\alpha$ от 0° до 90° профиля центрального тела будут изменяться нормальные скорости движения и зоны возмущения и связанные с нею характеристики. Эти изменения угла $\alpha$ центрального тела тоже связаны с исследованием величины сил, действующих на площадь сечения центрального тела ($\delta \cdot Ш$), и исследованием обтекаемости тела.

В зоне возмущения участка потока III-IV на профиль центрального тела будут действовать статические и динамические силы давления поступательного и нормального движения. Рассмотрим их действие.

Поступательные динамические силы давления направлены от поверхности центрального тела, поэтому они не оказывают на него силового воздействия.

Нормальные динамические силы давления направлены перпендикулярно оси потока. Тем самым они уравновешивают друг друга и не оказывают осевого воздействия на центральное тело.

Статические силы давления оказывают силовое воздействие на профиль центрального тела. Их величина зависит от величины динамических сил давления.

Изменение угла $\alpha$ профиля центрального тела изменяет величину нормальных динамических сил давления, которые, в свою очередь, влияют на величину статических сил давления.

При угле $\alpha$, стремящемся к нулю, нормальные динамические силы давления тоже стремятся к нулю. В этом случае величина статических сил давления определяется только поступательными динамическими силами давления зоны возмущения. Поэтому по отношению к нормальным динамическим силам давления статические силы давления будут иметь максимальную величину.

При угле $\alpha$, стремящемся к 90°, нормальные динамические силы будут стремиться к бесконечности. В этом случае величина статических сил давления будет зависеть от величины нормальных динамических сил давления зоны возмущения участка потока III-IV. В зависимости от этих сил мы здесь получим минимальное значение статических сил давления. Следовательно, варьируя углом $\alpha$ профиля центрального тела на участке потока III-IV, мы изменяем в определённом диапазоне величину нормальных динамических сил давления и величину статических сил давления.

Обтекаемость профиля центрального тела на участке потока III-IV нарушается при стремлении угла $\alpha$ к 90°. В этом случае мы можем нарушить распределение полной энергии потока между потенциальной и кинетической. Максимальные нормальные скорости зоны возмущения в этом случае будут поддерживаться за счет потерь полной энергии потока.

ПРИМЕР № 4

В данном примере требуется определить силы, действующие на крыло, которое перемещается в среде неподвижной жидкости или газа.

Под термином «крыло» принято понимать твёрдое тело определённого асимметричного профиля, которое при движении в среде неподвижного газа или жидкости создает определённую подъёмную силу. Среда здесь определяется как некоторый объём жидкости или газа, который находится в состоянии покоя под действием векторного или скалярного силового поля. Следовательно, жидкости или газы в этом случае обладают определённым запасом потенциальной энергии и имеют определённые статические силы давления.

Дальше полагаем, что крыло движется с определённой постоянной скоростью в среде жидкости или газа. Покажем его движение на рисунке 30. Передний профиль крыла образован окружностью с радиусом $r$, задний – треугольником с углом задней кромки крыла равной углу $\alpha$.

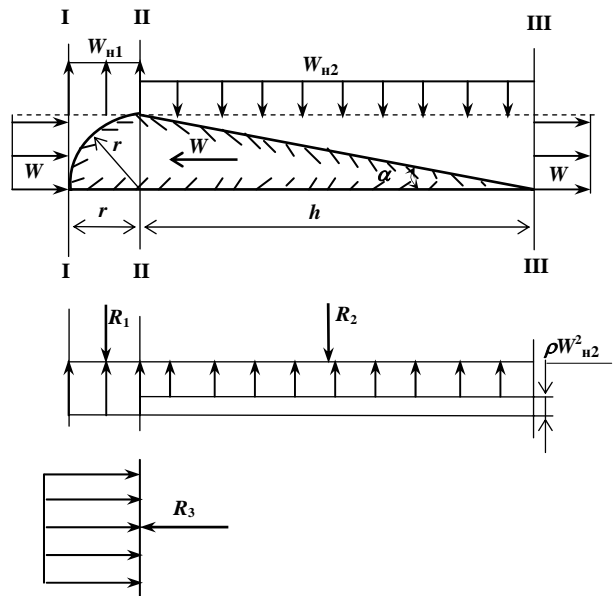

Рис. 30

Крыло движется с постоянной скоростью $W$. В свою очередь, жидкость или газ будут набегать на крыло тоже с постоянной скоростью равной скорости движения этого крыла. В тоже время жидкости и газы при движении относительно профиля крыла испытывают пространственное изменение. Поэтому для решения этой задачи воспользуемся зависимостями установившегося и плоского установившегося видов движения, как это сделано в примере № 3. В соответствии с эти примером определим границы зон возмущения для профиля крыла, как показано на рисунке 30. Затем переходим к определению зависимостей для данного профиля.

За полную энергию набегающего установившегося потока принимаем удельную энергию среды $U_{ср}$.

Составляем уравнение движения зоны возмущения, которая ограничивается плоскостями I и II. В плоскости I для поступательной плоскости исследования оно будет иметь вид:

$$M_{п1} = \rho W r. \tag{17}$$

В уравнении (17) для площади сечения мы принимаем длину крыла за единицу длины. Поэтому площадь здесь записана произведением единицы длины на радиус $r$. Для других зависимостей мы тоже будем относить площадь к единице длины.

В уравнении (17) нам известна плотность $\rho$. Как для жидкостей, так и для газов она равна плотности неподвижной среды. Также скорость $W$ – она равна скорости движения крыла, но имеет противоположное направление. Следовательно, по уравнению движения (17) мы можем определить расход массы в единицу времени для зоны возмущения I – II, так как этот расход будет равен нормальному расходу массы зоны возмущения, то есть $M_{\text{п}}{=}M_{\text{н}}$.

Запишем уравнение движения для нормального движения на площади границы зоны возмущения. Оно будет иметь вид:

$$M_{\text{н1}}{=}\rho W_{\text{н1}} r. \tag{18}$$

Из уравнения движения (18) мы можем определить нормальную скорость движения зоны возмущения I – II, так как в этом уравнении нам известны все величины, кроме нормальной скорости. В данном конкретном случае она будет равна поступательной скорости зоны возмущения, в связи с тем, что площади нормального и поступательного потоков зоны возмущения равны между собой.

Уравнение движения для объёма потока участка I – II составлять не требуется. Так как среда неподвижна, и объём её есть бесконечно большой объём. Мы получили полный комплект уравнений движения для зоны возмущения участка крыла I-II.

Составляем уравнения сил для зоны возмущения участка профиля крыла I – II.

Для поступательного движения зоны возмущения участка I – II оно будет иметь вид:

$$P_{\text{пр.п}}{=}P_{\text{ст.п}}{+}\rho W^2. \tag{19}$$

Для нормального движения зоны возмущения участка I – II оно будет иметь вид:

$$P_{\text{пр.н}}{=}P_{\text{ст.н}}{-}\rho W_{\text{н}}^2. \tag{20}$$

Затем по уравнению энергии установившегося вида движения или по уравнению Бернулли определяем величину статических сил давления для нормального и поступательного потоков участка I – II. Полная энергия потока нам известна – это потенциальная энергия неподвижной среды. Кинетическую энергию мы можем вычислить, так как нам известна плотность и скорость нормального и поступательного потоков зоны возмущения. В нашем конкретном примере нормальные и поступательные статические силы давления будут равны друг другу. Вычислив статические силы давления, мы сможем определить все действующие силы на участке I – II.

Теперь нам остается определить силы, действующие на профиль крыла участка II – III.

Полная энергия установившегося потока будет равна потенциальной энергии среды $U_{\text{ср}}$.

*Уравнение движения поступательной плоскости* III для площади зоны возмущения II – III будет иметь вид:

$$M_{\text{п1}}{=}\rho W r. \tag{21}$$

Это уравнение по всем величинам равно уравнению (17).

*Уравнение* движения для нормального движения на границе зоны возмущения участка II – III будет иметь вид:

$$M_{\text{н2}}{=}\rho W_{\text{н2}} h. \tag{22}$$

Из этого уравнения мы сможем определить величину нормальной скорости движения в зоне возмущения.

*Уравнение сил поступательного* движения зоны возмущения участка II – III будет иметь вид:

$$P_{\text{пр.п}}{=}P_{\text{ст.п}}{-}\rho W^2. \tag{23}$$

*Уравнение сил нормального* движения зоны возмущения участка II – III будет иметь вид:

$$P_{\text{пр.н}}{=}P_{\text{ст.н}}{-}\rho W_{\text{н2}}^2. \tag{24}$$

Затем по уравнению энергии установившегося вида движения делаем распределение полной энергии потока на потенциальную и кинетическую части для поступательного и нормального движения и определяем статические силы давлений этих движений на участке II – III.

Вычислив статические силы давления зоны возмущения участка II – III, мы получим все количественные величины для всего профиля крыла. Теперь мы сможем составить условие равновесия для движущегося крыла.

## Подъёмная сила

Подъёмные силы крыла действуют перпендикулярно направлению его движения. На участке потока I – II со стороны зоны возмущения действуют статические и динамические силы давления. Нормальные динамические силы давления этого участка направлены от профиля крыла. Поэтому они не оказывают на него воздействия. Поступательные динамические силы давления этого участка действуют на профиль крыла в направлении его движения. Поэтому они тоже не влияют на подъёмную силу профиля.

Со стороны зоны возмущения участка I – II на профиль крыла будут действовать только статические силы давления, которые для поступательного и нормального движения в нашем примере одинаковы.

Со стороны невозмущённой среды на площадь участка I – II профиля крыла будут действовать статические силы давления, которые соответствуют полной энергии установившегося потока или потенциальной энергии среды. По величине они больше статических сил давления зоны возмущения участка I – II. Если обозначить подъёмную силу крыла участка I – II через $R_1$, то она будет равна:

$$R_1 = (P_{ст.ср} - P_{ст1}) \cdot r. \tag{25}$$

В зоне участка II – III на профиль крыла тоже будут действовать статические и динамические силы давления.

Со стороны зоны возмущения участка II – III действует поступательные динамические силы давления, которые действуют в противоположную скорости движения крыла сторону, и действие их направлено от профиля. Поэтому они не оказывают силового воздействия на него.

Нормальные динамические силы давления участка II – III действуют непосредственно на профиль крыла. Их направление совпадает с направлением действия подъёмной силы крыла.

Статические силы участка II – III в зоне возмущения тоже действуют на профиль крыла. Их действие мы тоже должны учесть при определении подъёмной силы крыла. Мы знаем, что, в зависимости от величины нормальных и поступательных сил давления, статические силы могут быть разными. В данном случае статическое давление принимается непосредственно в зависимости от большей величины нормальных или поступательных динамических сил давления. В нашем случае поступательные динамические силы имеют большую величину, чем нормальные динамические силы давления. Поэтому для расчёта подъёмной силы крыла мы возьмем статические силы давления поступательного движения зоны возмущения участка II – III.

Со стороны невозмущенного потока будут действовать статические силы давления, величина которых соответствует величине полной энергии установившегося потока или потенциальной энергии среды.

Обозначим подъёмную силу участка профиля через $R_2$, тогда ее величина будет равна разности статического давления невозмущенной среды с нормальными динамическими силами возмущенного участка II – III и поступательными статическими силами того же участка профиля. Запишем величину подъёмной силы:

$$R_2 = (P_{ст.ср} - \rho W_{н2}^2 - P_{ст.п}) \cdot h. \tag{26}$$

Мы получили подъёмную силу для всего профиля крыла, изображенного на рисунке 30.

Зная, какие характеристики влияют на величину подъёмной силы крыла, можно сравнительно просто управлять ею.

## Лобовое сопротивление крыла

Нам остается определить лобовое сопротивление крыла. Опять начнем с распределения сил давления на участке профиля I – II. Силы, определяющие лобовое сопротивление крыла, действуют в направлении его движения. Такими силами в зоне возмущения профиля участка I – II являются поступательные динамические и статические силы давления. В зоне возмущения участка II – III поступательные динамические силы давления направлены от профиля крыла. Поэтому они не оказывают на него силового воздействия, и мы их не будем учитывать при определении лобового сопротивления. Остаются только поступательные статические силы давления этого участка. Обозначим силу сопротивления через $R_3$. Она будет равна сумме поступательных динамических и статических сил давления за вычетом поступательных сил давления зоны возмущения участка профиля II – III. Запишем эту силу лобового сопротивления:

$$R_3 = (P_{ст.п1} + \rho W^2 - P_{ст.п2}) \cdot r. \tag{27}$$

В нашем примере поступательные статические силы участка профиля I – II и II – III равны между собой. Тогда сила лобового сопротивления будет равна только поступательным динамическим силам давления зоны возмущения участка I – II, то есть:

$$R_3 = \rho W_{\text{п}1}^2 \cdot r. \tag{28}$$

Мы определили подъёмную силу и лобовое сопротивление крыла в зависимости от его профиля.

Отметим, что при расчёте подъёмной силы крыла в реальных вязких жидкостях и газах необходимо будет нормальные динамические силы давления в уравнении (26) умножить на коэффициент кинематической вязкости μ, а при расчёте сил лобового сопротивления необходимо будет умножить поступательные силы давления в уравнениях (27) и (28) на динамический коэффициент вязкости ν.

В примере № 4 мы рассмотрели влияние условий практического использования жидкостей и газов в их качестве неподвижной среды при движении профиля крыла. Как видим, и эти условия требуют решения своих проблем.

ПРИМЕР № 5

В данном примере нам необходимо определить состояние набегающего потока жидкости или газа по энергетическому состоянию при движении твёрдого тела в неподвижной среде. При условии, что твёрдое тело в каждый новый рассматриваемый момент времени будет двигаться с большей скоростью, чем в предыдущий момент движения.

В общем, сущность этого примера будет состоять в том, чтобы определить изменение энергетического состояния набегающего потока в зависимости от скорости твёрдого тела.

Потенциальная механическая энергия естественной среды, то есть воды или воздуха, определяется природными условиями, которые создает для них наша планета с помощью силовых полей. Величину ее мы можем получить при помощи соответствующих замеров.

По отношению к движущемуся предмету среда образует поток установившегося вида движения. Потенциальная энергия среды является полной энергией этого потока. Движущийся предмет, согласно уравнению энергии установившегося вида движения или уравнению Бернулли, производит перераспределение полной энергии потока или потенциальной энергии среды между потенциальной и кинетической энергиями этого образовавшегося потока.

Из условий движения установившегося потока реальных невязких жидкостей и газов в трубопроводах нам известны предельные условия распределения полной энергии потока между кинетической и потенциальной. Они выражаются в том, что кинетическая и потенциальная энергия потока по величине достигают, каждая в отдельности, половины полной энергии потока ($U_{\text{к}} = U_{\text{п}} = \dfrac{U}{2}$). Тогда скорости движения жидкостей и газов становятся определённой постоянной величиной, которую можно увеличить лишь при увеличении полной энергии установившегося потока. Для газа скорость движения при предельных условиях распределения достигает скорости звука и называется критической скоростью.

В тоже время, мы знаем, что различные предметы могут двигаться в среде жидкостей и газов со скоростью, которая превышает, например, скорость звука для данной среды. По этой причине энергетическое состояние образующегося набегающего потока должно быть иным, чем это записано уравнением механической энергии установившегося вида движения или уравнением Бернулли. В этом примере требуется определить это новое энергетическое состояние.

Рассматриваем движение предмета со сверхзвуковой скоростью в среде газа. Что здесь нам известно из практики? Только величина скорости движения и локальное качественное изменение среды вокруг движущегося предмета, которое, например, выражается в повышенных температурах. Если мы теперь попытаемся записать для него уравнение распределения энергии в обычном виде, то нам этого не удастся сделать, поэтому мы должны найти иное распределение энергии набегающего потока.

Как мы выше установили, что при движении предмета в среде газа со сверхзвуковой скоростью образуется установившийся поток газа с механической и термодинамической энергией. В разделе о реальных жидкостях и газах мы установили, что полная энергия такого потока записывается уравнением энергии (VII.8). Выпишем его еще раз:

$$U_{\text{д}} = U_{\text{п}} + \int\limits_0^{P_2} dP \pm \int\limits_1^{V_2} P dV. \tag{VII.8}$$

Согласно этому уравнению и положениям, связанным с ним, для установившегося потока следует, что при изменении термодинамических свойств массы газа или жидкости механическая энергия потока остается неизменной, а прирост энергии происходит за счет изменения термодинамических свойств массы жидкости и газа. В этом уравнении механическая и термодинамическая энергии являются отдельными самостоятельными членами. В этом случае распределение полной механической энергии между потенциальной и кинетической энергией достигает своего крайнего предела. То есть для невязких жидкостей и газов величины кинетической и потенциальной энергий потока становятся равны половине величины полной механической энергии потока. Из этого условия следует, что верхний предел определённого интеграла кинетической энергии (VII.8) достигает своей предельной величины и становится равным максимальному динамическому значению сил давления $P_{max}$. Мы знаем, как можно определить величину этого давления. Она равна величине статического давления для полной механической энергии потока. В нашем случае она равна статическому давлению для неподвижной среды с соответствующей величиной потенциальной энергии.

Член уравнения (VII.8), учитывающий изменение термодинамической энергии, будет в нашем случае положительной величиной, так как у нас происходит прирост энергии. Теперь перепишем уравнение (VII.8) с учетом этих изменений:

$$U_{д} = U_{п} + \int_{0}^{P_{max}} dP + \int_{1}^{V_2} P dV. \tag{29}$$

Мы получили уравнение энергии установившегося потока газа, образующегося при движении предмета со сверхзвуковой скоростью в среде газа.

Теперь нам необходимо выразить энергию через характеристики потока. Для чего сначала запишем кинетическую энергию потока через его характеристики. Тогда получим:

$$U_{к\,max} = \frac{\rho W^2}{2} = \text{const}. \tag{30}$$

Рассмотрим все характеристики уравнения кинетической энергии (30). Максимальная кинетическая энергия $U_{к\,.max}$ есть величина постоянная в данном примере и равная половине потенциальной энергии среды или половине полной энергии установившегося потока ($U/2$). Следовательно, ее величину мы можем определить.

Скорость потока в этом случае можно определить через скорость движения предмета в среде. Скорость движения предмета в среде и скорость движения газа в возмущенной зоне среды будут равны друг другу. Это определение относится, прежде всего, к поступательной скорости движения. Следовательно, скорость движения газа мы тоже можем определить.

В этом уравнении остается неизвестным для нас только плотность газа $\rho$. Используя уравнение (30), мы можем определить ее величину в каждом конкретном случае. Проанализировав изменение плотности газа в зависимости от скорости, мы придем к выводу, что с увеличением скорости при сверхзвуковом движении предмета плотность среды уменьшается. При этом зависимость здесь квадратичная. Мы получили кинетическую энергию потока, выраженную через его характеристики.

Чтобы определить действительную энергию потока $U_{д}$, мы должны знать величину термодинамической энергии в уравнении (29). Ее величину мы можем определить только по изменению плотности. Но для этого надо знать соответствующую зависимость термодинамического процесса. С этим вопросом обратимся к термодинамике. Согласно ее положениям в данном случае мы должны были бы применить зависимость адиабатического процесса, но его зависимость неприменима для нашего случая. Так как с ростом энергии уменьшается плотность среды. Пожалуй, в данном случае нам подойдет зависимость изобарного процесса, но этот процесс рассматривает изменение термодинамических характеристик газа при постоянном давлении, а не при постоянной величине механической энергии. Так что и здесь мы можем попасть в ловушку. Следовательно, термодинамика не дает четкого ответа в этом вопросе. Поэтому для дальнейших рассуждений принимаем, что изменение плотности газа в потоке происходит по некоторой зависимости политропного процесса. Будем считать, что нам известна зависимость термодинамических изменений состояния массы газа.

Тогда величину термодинамической энергии мы определим следующим образом. Пределы изменения плотности газа мы определим по уравнению кинетической энергии (30). Затем с помощью уравнения термодинамического процесса найдем пределы изменения единицы объёма ($V_2$) для определённого интеграла уравнения (29), который выражает количество термодинамической энергии. После этого заменим в этом интеграле давление $P$ с помощью зависимости термодинамического процесса и вычислим полную величину термодинамической энергии установившегося потока газа. Следовательно, теперь мы можем определить количество термодинамической энергии для установившегося потока газа, образованного движущимся предметом, скорость которого превышает скорость звука неподвижного газа. Сложив потенциальную энергию среды и термодинамическую энергию, мы получим действительную энергию потока газа.

При движении предметов с еще большими скоростями, то есть со скоростями, которые известны нам из практики, никаких других качественных изменений среды не наблюдалось. Поэтому не будем дальше рассматривать движение предметов в среде газа с еще большими скоростями. Хотя скачок уплотнения оказался скачком «разуплотнения» при рассмотрении движения предмета в среде со сверхзвуковыми скоростями, но мы все равно получили необходимые зависимости, которые характеризуют качественное изменение в газах в зависимости от их термодинамических свойств. Следовательно, скачок уплотнения мы должны понимать как переход массы газа от одних термодинамических свойств к другим при определённых условиях движения.

Теперь рассмотрим движение предмета с постоянной скоростью в среде жидкостей. При движении предметов у поверхностей водных бассейнов нашей планеты до глубины не более 5 метров уже при скорости 10-15 м/сек происходит предельное распределение полной энергии набегающего потока или потенциальной энергии среды между потенциальной и кинетической, то есть, без учета вязкости, величина потенциальной и кинетической энергии потока достигает половины полной энергии потока. Поэтому, при движении предмета в бассейнах нашей планеты со скоростями больше 15 м/сек, мы должны будем учитывать в уравнении полной энергии потока еще и термодинамическую энергию. Следовательно, при скорости движения предмета 10-15 м/сек происходит скачек уплотнения, аналогичный скачку уплотнения газов. С этим скачком у людей не было никаких неприятностей. Пожалуй, по этой причине он остался незамеченным. В то же время при определении действительной энергии набегающего потока мы обязаны воспользоваться уравнением (29). Порядок определения этой энергии такой же, как и для газов. Полная механическая энергия набегающего потока берется как потенциальная энергия среды. Плотность жидкости в возмущенной зоне определяется по уравнению (30). Термодинамическая энергия определяется по изменению плотности в соответствии с зависимостью термодинамического процесса. Затем определяется действительная энергия набегающего потока как сумма полной механической энергии набегающего потока и термодинамической энергии.

Отметим, что для жидкости в термодинамике нет [описания] каких-либо процессов, которые бы через зависимости выражали ее термодинамические свойства. Так как жидкость считается несжимаемой. Просто переход тепловой энергии в механическую определяют при помощи соответствующего эквивалента. Будем надеяться, что она может быть «разжимаемой».

Считаем, что мы получили все необходимые зависимости для первого скачка термодинамических свойств жидкости.

С увеличением постоянной скорости движения предмета в среде жидкости растет и температура жидкости в набегающем потоке. Наконец, скорость достигнет такой величины, когда термодинамическое состояние жидкости станет неустойчивым и жидкость начнет переходить в пар. В обычной жизни это состояние называют кипением жидкости. Для жидкости набегающего потока подобное состояние будем называть кавитацией. Для нее это будет второй скачек изменения чисто термодинамических свойств. В водных бассейнах нашей планеты для набегающего потока кавитация наступает при движении предмета со скоростью порядка 40-45 м/сек.

Кавитацию можно назвать своеобразным процессом кипения жидкости в набегающем потоке, который образован движущимся в среде жидкости предметом. С помощью энергетических зависимостей можно определить только границы для каждой жидкости, в пределах которых существует кавитация.

Если дальше рассматривать увеличение постоянной скорости предмета, то мы увидим, что при достижении определённой величины скорости, кавитация в набегающем потоке прекращается и жидкость поступает в поток сразу в газообразном или парообразном состоянии. Дальнейшее увеличение скорости движущегося предмета надо рассматривать как увеличение его скорости при движении в среде газа.

Отметим, что движение предмета в жидкостях и газах рассматривалось как движение твёрдого тела абсолютно обтекаемой формы. Для необтекаемых тел, например, таких как плоскость, соответствующие скачки наступают при меньших скоростях движения, чем для обтекаемых.

Величину изменения плотности в набегающем потоке жидкости и газа можно контролировать путем измерения объёма или толщины пристеночного слоя, так как его величина определяет геометрические размеры набегающего потока.

Из этого примера следует, что набегающий установившийся поток жидкостей или газов воспринимает механическую энергию движущегося тела в виде распределения полной механической энергии среды между потенциальной и кинетической, а также за счет изменения термодинамического состояния жидкостей и газов в этом потоке.

В примере № 5 условия практического использования жидкостей и газов определяют ряд проблем, которые касаются не только механики жидкости и газа, но и термодинамики.

ПРИМЕР № 6

Даны два установившихся потока жидкости или газа с полной энергией потока $U$ и с постоянным расходом массы в единицу времени $M$. Потоки вытекают в окружающую среду с потенциальной энергией $U_{ср}$. Требуется определить размеры и профиль рабочей полости турбины, которая используется для преобразования энергии этих потоков в механическую работу.

Турбины относятся к лопастным машинам. В соответствии с положениями механики жидкости и газа принцип их работы основан на плоском установившемся виде движения жидкостей и газов. В зависимости от характера движения жидкостей и газов в рабочей полости все турбины можно разделить на два типа: радиальные и осевые. В свою очередь рабочая полость этих турбин делится на два участка. Границы участков определяются в соответствии с энергетическим состоянием жидкостей и газов в объёме плоского установившегося потока. Энергетическое состояние жидкостей и газов показано на рис. 26, гл.V. Согласно этому рисунку первый участок располагается от точки $A_2$ до точки $A_0$. Для турбины он определяется как участок от входа до критического сечения. На этом участке происходит преобразование части потенциальной энергии потока в работу и увеличение скорости потока от входных скоростей до максимально возможных скоростей в критическом сечении.

Второй участок располагается от точки $A_0$ до точки $A_3$. Для турбины он определяется как участок от критического до выходного сечения. На этом участке происходит преобразование части кинетической энергии в работу. Если это газовая турбина, то на этом же участке дополнительно происходит преобразование энергии сжатия в работу. Одновременно происходит снижение скоростей потока от максимально возможных до выходных скоростей. По этому участку различаются газовые и жидкостные турбины.

Турбины, полость которых состоит из двух участков, дают возможность целиком использовать полную энергию установившегося потока жидкости или газа.

При практическом использовании радиальных турбин для конкретных условий применения не всегда возможно сделать оба ее участка радиальными. В этом случае, например, можно будет сделать первый участок турбины радиальным, а второй – осевым. В нашем примере возьмем турбину именно такого типа. Так как она дает возможность рассмотреть одновременно конструктивные особенности радиальных и осевых турбин.

Зависимости, определяющие движение, силовое воздействие и распределение энергии в полости турбины, относятся к плоскому установившемуся виду движения идеальной жидкости. Для реальных жидкостей и газов их необходимо будет дополнить коэффициентами вязкости и зависимостями связанными с сжимаемостью.

Теперь перейдем к определению рабочей полости турбины. Покажем ее на рис. 31.

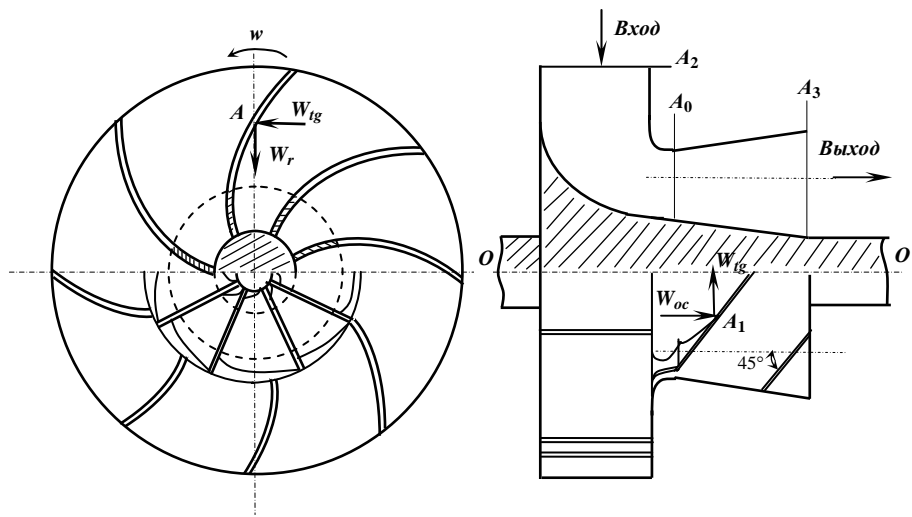

Рис. 31

1. Определяем площади сечения турбины на входе (точка $A_2$), на выходе (точка $A_3$) и в критическом сечении (точка $A_0$). Для этого воспользуемся зависимостями установившегося вида движения. На входе и на выходе площади сечения определяются из условия минимальных скоростей движения потока. Площадь критического сечения определяется из условия максимальных скоростей движения потока. Максимум скорости ограничивается предельным распределением полной энергии установившегося потока между потенциальной и кинетической по уравнению энергии установившегося вида движения. Для реальных жидкостей и газов эти площади определяются с учетом кинематического коэффициента вязкости.

2. Исходя из полученных сечений, определяем предварительные габариты рабочей полости турбины.

3. Размещаем лопасти турбины в ее рабочем объёме. Лопасти турбины размещаются по линиям тока или поверхностям тока.

Для первого участка турбины, который ограничивается входным и критическим сечениями, линиями тока являются логарифмические спирали, лучи которых пересекаются касательными линиями под углом 45°. В радиальном участке лопасти располагаются непосредственно по логарифмической спирали. Затем они переходят в лопасти осевого участка, плоскость которых располагается под углом 45° к осевой скорости потока и перпендикулярно к оси турбины $O$-$O$ (рис. 31). Длина их ограничивается плоскостью критического сечения и плоскостью выхода. Профиль или толщина лопастей турбины должна быть такой, чтобы расчётные площади сечения, входа, выхода и критического сечения, полученные в п.1 данного примера, оставались бы неизменными. Тогда действительные величины площади сечения турбины будут равны расчётным, к которым прибавляются площади сечения лопастей.

Руководствуясь этими, сравнительно простыми правилами и положениями, мы получили нужную для наших целей турбину с полным использованием энергии установившегося потока жидкости или газа. Теперь мы должны уяснить себе эти простые правила.

В первую очередь обратим внимание на условия выбора и расположения лопастей турбины. Основными условиями для них являются условия прочности лопастей и условия сохранения рабочей формы.

Условия прочности, с одной стороны, определяются силами механики твёрдого тела, к которым относятся центробежные силы, с другой стороны, – силами механики жидкости и газа. На первом участке такими силами являются тангенциальные силы давления, которые записаны уравнениями (IV.12). Тангенциальные силы давления состоят из статических и динамических сил давления. Тангенциальные статические силы действуют с двух противоположных сторон на лопасти турбины. Поэтому они уравновешивают друг друга и не оказывают силового воздействия на них. Следовательно, только тангенциальные динамические силы давления оказывают силовое воздействие на лопасти турбины. Согласно уравнению (IV.12), величина их на первом участке, от входа в турбину до критического сечения, непрерывно увеличивается. В то же время действительная их величина зависит еще от числа лопастей турбины. Чтобы выявить их влияние, поступим следующим образом: сначала запишем суммарную или общую тангенциальную динамическую силу давления для всего плоского установившегося потока как произведение расхода массы в единицу времени на скорость. Для чего принимаем расход массы как расход массы $M_{tg}$ всего потока в тангенциальном направлении, а скорость как некоторую среднюю величину $W_{tg\,ср}$ для потока первого участка турбины. Тогда получим:

$$R_{tg} = M_{tg}W_{tg\,ср}. \qquad (31)$$

Эта тангенциальная сила $R_{tg}$ не зависит от числа лопастей турбины и является постоянной величиной. Если бы эта величина зависела от числа лопастей турбины, то в зависимости от их числа менялся бы крутящий момент турбины. Чего не происходит в действительности.

Чтобы получить величину тангенциальной силы, которая приходится на долю одной лопасти, мы обязаны разделить эту силу на число лопастей. Обозначим число лопастей через $n$, тогда получим:

$$\frac{R_{tg}}{n} = \frac{M_{tg}W_{tg\,ср}}{n}. \qquad (32)$$

Рассмотрим уравнение (32) по смысловому значению, которое определяет размещение лопастей в объёме турбины. Если мы разделим скорость на число лопастей $n$, то тем самым уменьшим ее величину, а это уменьшение приведет к изменению динамических сил давления для всего потока в целом и непосредственно самого потока. Поэтому делать этого нельзя. Следовательно, на число лопастей можно делить только расход массы в единицу времени. Это значит, что лопасти турбины делят общий поток на ряд потоков, которые в сумме составляют объём расчётного потока без изменения его характеристик.

Уравнения (31) и (32) являются приближенными уравнениями, которыми мы воспользовались для выявления числа лопастей турбины. Поэтому действительную величину тангенциальных сил давления мы должны определять по уравнению (IV.12). Затем полученную величину разделить на число лопастей турбины $n$ и только после этого получим правильную величину тангенциальных динамических сил давления, действующих на лопасти турбины. Тогда уравнение (IV.12) с учетом числа лопастей примет вид:

$$P_{tg\,пр.д} = P_{tg\,ст} + \frac{1}{n}\rho W_{tg}^2. \qquad (33)$$

С точки зрения условий прочности, увеличение числа лопастей турбины приводит к уменьшению величины тангенциальных динамических сил давления. Таким образом, мы определились относительно условий прочности для радиальной ступени турбины.

Отметим, что здесь необходимо обратить внимание на различие действий по отношению к силам между положениями механики твёрдого тела и механики жидкости и газа.

Теперь рассмотрим условие сохранения расчётной формы рабочей полости турбины. В данном примере расчётная форма рабочей полости радиальной турбины соответствует объёму плоского установившегося потока, то есть радиальная полость турбины как бы копирует этот объём. Анализ влияния этих условий делается по уравнению энергии установившегося вида движения или уравнения Бернулли. Здесь могут быть два случая.

В первом случае рабочая полость турбины в силу каких-либо причин будет выполнена больше или меньше расчётной. Тогда при меньших размерах полости через турбину пройдет меньший расход массы, чем это предусмотрено в условиях этого примера, а при больших размерах полости через эту турбину должен пройти больший расход массы, чем это предусмотрено в условиях этого примера. Оба эти случая потребуют изменения полной энергии заданного установившегося потока.

Во втором случае профиль рабочей полости турбины в силу каких-либо причин будет выполнен с отклонением от расчётного. Например, лопасти турбины будут выполнены по логарифмической спирали с большим или меньшим, чем 45°, углом пересечения ее лучей с касательными, Это отклонение профиля лопастей приведет к перераспределению площадей между радиальными и тангенциальными расходами массы в единицу времени, которое потребует либо увеличение, либо уменьшение тангенциальной скорости движения при расчётном общем объёме потока. По этой причине подобные нарушения будут происходить за счет потерь полной энергии заданного установившегося потока. Величину этих потерь можно определить только экспериментальным путем.

Мы рассмотрели влияние условия сохранения расчётной формы рабочей полости радиальной ступени турбины. Отметим, что расположение лопастей турбины определяет также направление ее вращения по часовой стрелке или против нее.

Рассмотрим теперь осевую ступень турбины. Для этого нам придётся вспомнить основные положения плоского установившегося вида движения. Согласно положениям плоского установившегося вида движения объём этого потока является цилиндром, в каждой точке которого жидкость движется во взаимно перпендикулярных направлениях. Определяющим расходом потока является радиальный расход массы в единицу времени. Величина скорости этого расхода определяется площадью сечения потока. Площадь сечения потока представляет собой цилиндрическую поверхность. Тангенциальный расход и скорость определяются радиальным расходом массы. Линии тока объединяют все это движение. Как видим, все его зависимости только в сумме определяют форму плоского установившегося потока, а по раздельности они допускают другие варианты и комбинации своего применения. Выше мы их применяли при разборе установившегося вида движения. Теперь с соответствующими изменениями применим зависимости радиального потока к осевому.

Заменим в осевой ступени турбины радиальный расход массы на осевой $M_{ос}$. Этот расход является определяющим для осевого потока. Затем, расположив лопасти перпендикулярно к оси потока и под углом 45° к осевой скорости движения $W_{ос}$, мы тем самым определим [другое] направление радиальной скорости движения $W_r$ и радиального массы $M_r$. Для этого полости площади сечения находятся в плоскостях, которые располагаются перпендикулярно оси потока $O$-$O$. Величина осевой скорости определяется площадью сечения осевого потока. Вот все отличительные особенности осевого потока от радиального. Поэтому все зависимости радиального потока, которые определяют скорости движения и силы давления, пригодны и для осевого потока. В этом случае мы только должны для соответствующих зависимостей заменить знаки, обозначающие радиальное движение, на знаки, обозначающие осевое движение, а тангенциальные зависимости должны остаться без изменения. Если бы, например, в нашем примере радиальная ступень турбины была бы выполнена осевой, то мы все зависимости и положения этой ступени должны были бы оставить без изменений, за исключением линии тока, где мы должны были бы заменить логарифмическую спираль на плоскость тока, которая расположена перпендикулярно к оси потока и под углом 45° к осевой скорости движения. Будем считать, что мы рассмотрели первый участок турбины, выполненный в осевом варианте.

Второй участок турбины, где происходит преобразование кинетической энергии в работу, выполнен осевым. Для газов на этой же ступени одновременно происходит преобразование энергии сжатия в работу. Поэтому увеличение площади сечения потока должно быть выполнено с учетом расширения газа. Только в этом заключается основное различие газовых и жидкостных турбин.

При преобразовании кинетической энергии в работу на лопасти турбины будут оказывать силовое воздействие осевые динамические силы давления. Если бы эта ступень была радиальной, то мы бы назвали эти силы радиальными. Зависимости, как для радиальных, так и для осевых сил давления, одинаковы, поэтому для их записи воспользуемся уравнением (IV.11). Мы его должны переделать с учетом числа лопастей турбины. Тогда действительная величина осевых динамических сил давления, действующих на единицу площади лопасти турбины, будет иметь вид:

$$P_{пр.д} = P_{ос.ст} + \frac{1}{n} \rho W_{ос}^2, \qquad (34)$$

где $n$ — число лопастей турбины.

Все остальные положения, которые мы применили для пояснения первого радиального участка, остаются верными и для данного участка турбины. Получив разъяснение для ступеней турбины, рассмотрим ее теперь как одно целое.

Силы давления жидкостей и газов действуют в неподвижных плоскостях, поэтому величина их действия не зависит от числа оборотов турбины. Это значит, что величина крутящего момента турбины есть величина постоянная и не зависит от того, вращается ли она или закреплена неподвижно. Отсюда следует, что все необходимые замеры характеристик потока турбины можно делать в неподвижном для нее состоянии, то есть закрепив вал турбины и пропустив через ее рабочую полость заданный поток.

Определим коэффициент полезного действия турбины. Он определяется как отношение действительной работы к располагаемой или теоретической. Запишем его:

$$\eta = \frac{L_д}{L_т}. \qquad (35)$$

Теоретическую работу в этом случае можно определить как разность между полной энергией потока и потенциальной энергией среды.

Для жидкости она будет равна:

$$L_т = U - U_{ср}, \qquad (36)$$

где $L_т$ – теоретическая работа, $U$ – полная энергия установившегося потока, $U_{ср}$ – потенциальная энергия окружающей среды.

Для газов она будет равна:

$$L_т = U - U_{ср} + \int_1^{V_{ср}} PdV. \qquad (37)$$

где $V_{ср}$ – объём газа, расширившегося до состояния [объёма] среды $V_{ср}$; $\int_1^{V_{ср}} PdV$ – величина энергии сжатия газового потока.

Действительная работа $L_д$ либо замеряется, либо определяется с учетом следующих потерь полной энергии установившегося потока. Здесь учитываются потери, связанные с вязкостью, термодинамическими условиями движения, отклонением рабочей полости турбины от расчётной и так далее. Подставив теоретическую и действительную работу в отношение (35), получим КПД турбины.

Будем считать, что мы пример № 6 выполнили.

Заметим, что полученная турбина, в отличие от современных турбин, не имеет соплового аппарата. Имеет скорости потока на входе и на выходе близкие к нулю. Из этого различия следует, что в современных турбинах совсем не использовали потенциальную и большую часть кинетической энергии установившегося потока, не говоря уже о потерях, связанных с профилированием. Судя по перечисленным недостаткам современных турбин, можно сказать, что их КПД не превышает КПД современных поршневых двигателей. По расчётам данного примера можно практически получить турбину с КПД близким к единице. Отсюда следует вывод, что нет смысла использовать какой-то другой роторный двигатель, который был бы экономичней и проще турбины.

В заключении еще раз вернемся к конструктивным особенностям радиальных и осевых турбин. Выше мы рассмотрели их по отношению к зависимостям плоского установившегося вида движения. Теперь рассмотрим по отношению к расположению рабочей полости турбины. Сущность такого расположения заключается в том, чтобы определить, возможно ли, сохранив расчётный объём этих турбин неизменным, разместить его на большем радиусе относительно оси турбины. В этом случае мы сможем сохранить величину динамических сил давления потока неизменными, при этом увеличив радиус их действия относительно оси турбины. Тогда мы сможем получить на турбине больший крутящий момент за счет увеличения радиуса.

Для радиальной турбины площадью сечения потока является цилиндрическая поверхность, ось которой совпадает с осью турбины. Для осевой турбины плоскость сечения потока лежит в плоскости, которая расположена перпендикулярно к оси турбины и к оси потока. Площадь сечения потока этих турбин определяет величину скорости и величину динамических сил давления, которые непосредственно оказывают силовое воздействие на ее лопасти. Поэтому, увеличивая площадь сечения потока, мы тем самым уменьшаем одновременно и скорости, и динамические силы потока. При этом уменьшении динамических сил давления пропорционально квадрату скорости. По этим причинам мы записали условие постоянства плоскостей сечения и объёмов рабочей полости турбины при увеличении радиуса потока.

В рамках этих условий попытаемся увеличить радиус потока. Для радиальной турбины это значит – увеличить одновременно радиус входа, радиус критического сечения и радиус выхода. Для осевой турбины это значит – увеличить радиус цилиндрической поверхности, относительно которого симметрично располагается объём ее рабочей полости. Поэтому в радиальной турбине площадь сечения потока находится в прямой зависимости от радиуса потока. Если мы теперь для нее начнем увеличивать радиус потока, то одновременно мы должны будем уменьшать ширину рабочей полости турбины, чтобы сохранить величину площадей сечения неизменными. Для этой турбины нам удастся незначительно увеличить радиус потока, так как ширина полости быстро станет такой, что через нее невозможно будет пропустить необходимый расход реальной вязкой жидкости. Используя эти сравнительно общие рассуждения для радиальной турбины, можно полагать, что в рамках вышеизложенных условий величину радиуса потока можно увеличить незначительно.

Если мы теперь начнем увеличивать радиус потока для осевой турбины, то мы должны будем уменьшать ширину ее рабочей полости пропорционально длине цилиндрической поверхности. Исходя из этих общих рассуждений, можно полагать, что величину радиуса потока осевой турбины можно увеличивать на сравнительно большую величину, чем увеличение радиуса радиальной турбины. Следовательно, при определении коэффициента полезного действия необходимо будет учитывать изменение радиуса потока турбин.

Пожалуй, на осевой турбине можно будет получить КПД больше единицы за счет увеличения радиуса потока. Это положение можно будет выяснить только после более детальных теоретических и экспериментальных исследований.

ПРИМЕР № 7

Требуется определить истечение жидкости и газа из круглого отверстия от действия статических сил давления.

Здесь имеются в виду отверстия, которые расположены в стенках объёмов, в которых жидкости и газы находятся в состоянии покоя под действием определённых силовых полей, или отверстия в стенках трубопроводов, где движется определённый установившийся поток. Покажем на рисунке 32 такое отверстие, с вытекающей из него жидкостью или газом. При вытекании жидкости или газа из отверстия образуются как бы два потока от действия разности сил давления окружающей среды и определённого объёма жидкости или газа.

Непосредственно у стенки объёма образуется поток плоского установившегося вида движения (рис. 32). В этом потоке жидкости и газы преобразуют потенциальную энергию в кинетическую. На внутреннюю границу плоского установившегося потока, который совпадает с диаметром отверстия в стенке (рис. 32) жидкости и газы поступают имея только кинетическую энергию. С внутренней границы потока к оси отверстия они движутся по линиям тока, которые имеют форму окружности с определённым радиусом. По такой форме линий тока нет потерь энергии. Таким образом происходит формирование границ струи жидкости или газа с внутренней стороны стенки относительно оси отверстия. Затем жидкость или газ поступает в отверстие стенки уже в виде сформировавшегося потока и дальше – во внешнюю среду.

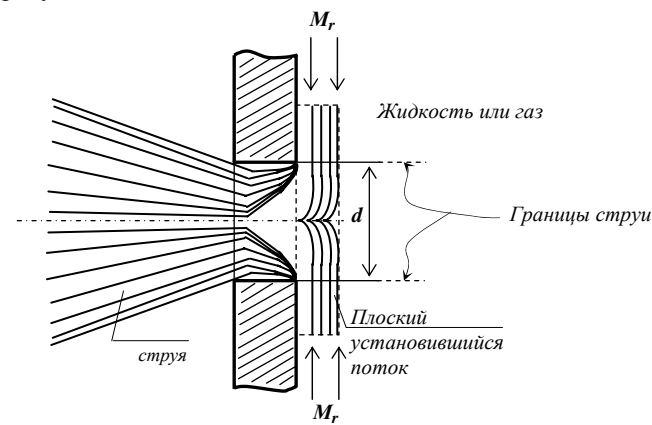

Рис. 32

Возьмем в качестве плоскости исследования струи внутреннюю поверхность стенки и запишем необходимые зависимости. Тогда уравнением движения струи будет уравнение движения установившегося вида движения. За площадь сечения струи принимаем площадь отверстия с диаметром $d$. Запишем его:

$$M_c = \rho W \frac{\pi d^2}{4}. \tag{38}$$

В этом сечении жидкости и газы не обладают статическим давлением, так как не имеют потенциальной энергии. По этой причине мы должны будем записать для струи только уравнение динамических сил давления. Оно будет иметь вид:

$$P_{\text{дин}} = \frac{4M_{\text{с}}}{\pi d^2} \cdot W. \qquad (39)$$

Воспользовавшись уравнениями (38) и (39), мы найдем все необходимые характеристики струи. Характеристики плоского установившегося потока определяются по зависимостям плоского установившегося вида движения. Поэтому мы их здесь не будем переписывать, так как каждый для себя сможет записать их самостоятельно.

В современной литературе, в отношении струи, предпочитают иметь зависимость расхода массы непосредственно от статических сил давления. Мы тоже постараемся получить подобную зависимость. Для чего заменим динамические силы давления $P_{\text{дин}}$ в уравнении (39) через разность статических сил давления среды $P_{\text{ст.ср}}$ и объёма $P_{\text{ст}}$. Затем заменим скорость $W$ в уравнении (38) по уравнению (39), тогда получим:

$$M_{\text{с}} = \frac{\pi d^2}{4} \sqrt{\rho\left(P_{\text{ст}} - P_{\text{ст.ср}}\right)}. \qquad (40)$$

Подобное уравнение называют уравнением расхода отверстия.

При определении расхода массы для реальных вязких жидкостей и газов мы обязаны умножить правую часть уравнения (40) на коэффициент кинематической вязкости $\mu$, тогда оно примет вид:

$$M_{\text{с}} = \mu \frac{\pi d^2}{4} \sqrt{\rho\left(P_{\text{ст}} - P_{\text{ст.ср}}\right)}. \qquad (41)$$

Уравнение (41) является уравнением расхода реальных вязких жидкостей и газов для отверстия круглой формы.

На практике часто применяют круглые отверстия с различной формой внутренней стороны стенки. Подобные отверстия могут иметь расход массы отличный от расхода массы отверстия с такой же площадью в плоской стенке. Подобное различие можно учесть с помощью коэффициента формы отверстия $\xi$. Запишем его как отношение расхода массы $M_{\text{ф}}$ отверстия с иной формой к расходу массы отверстия в плоской стенке $M_{\text{с}}$, получим:

$$\xi = \frac{M_{\text{ф}}}{M_{\text{с}}}. \qquad (42)$$

Этот коэффициент проще всего определить экспериментальным путем и свести его величины для характерных отверстий в таблицы. Тогда более общее уравнение расхода будет иметь вид:

$$M_{\text{с}} = \xi \mu \frac{\pi d^2}{4} \sqrt{\rho\left(P_{\text{ст}} - P_{\text{ст.ср}}\right)}. \qquad (43)$$

Уравнение (43) также пригодно для отверстий иных форм, например, квадратных, треугольных и так далее. Для них тоже необходимо будет ввести свой коэффициент формы отверстия.

Заметим, что в современной практике применяется уравнение расхода такого вида:

$$M_{\text{с}} = \mu \frac{\pi d^2}{4} \sqrt{2\rho\left(P_{\text{ст}} - P_{\text{ст.ср}}\right)}.$$

где $\mu$ называется коэффициентом расхода.

Уравнение в таком виде записано для невязкой жидкости. Поэтому сравним его с уравнением (40), которое тоже записано для невязкой жидкости. Оно отличается от уравнения (40) тем, что имеет коэффициент расхода $\mu$ и цифру 2 под корнем. Такое различие вызвано тем, что при выводе этого уравнения пользовались уравнением Бернулли, а не уравнением сил. Поэтому под корнем появилась двойка. Чтобы компенсировать эту ошибку, ввели в это же уравнение коэффициент расхода $\mu$, а его появление объяснили так называемым «сужением» струи.

Мы рассмотрели уже семь примеров, которые охватили состояние покоя жидкостей и газов, установившегося и плоского установившегося вида движения. Согласно порядку изложения следующий пример должен быть по расходному виду движения жидкостей и газов. В данной работе мы не будем его приводить, а сошлемся на пример № 3 работы [3][2], который является наиболее характерным для данного вида движения. Дальше перейдем к акустическому виду движения жидкости и газа.

---

ПРИМЕР № 8

Требуется рассчитать глушитель для глушения выхлопа газов на выходе из поршневых двигателей.

Непосредственно от конкретного двигателя для этой цели мы должны будем получить следующие данные:

массу газа на один выхлоп $m$, температуру $T_в$, давление $P_в$, плотность $\rho_в$ при начале выхлопа, величину времени выхлопа $t_в$, площадь выхлопного окна $F_в$.

Затем берут зависимости акустического вида движения и приступают к расчёту необходимого глушителя. Все очень просто, но не совсем. Прежде, чем приступить к расчёту, необходимо знать еще хотя бы конструктивную схему глушителя и принцип его работы. Эту схему и принцип работы можно получить непосредственно из знания природы и характера акустического вида движения жидкостей и газов. Для этого недостаточно знать зависимости этого вида движения. Необходимо сравнительно хорошо представлять себе картину движения при акустическом виде движения. Поэтому сначала постараемся уяснить ее себе. При этом нам придётся вспомнить свойства реальных жидкостей и газов.

Затем берем пластину источника возмущения конечной определённой площади $F$. Дальше полагаем, что эта пластина совершила поступательное движение из положение I в положение II сразу на всю длину своего хода $l$, а затем прекратила свое движение. Покажем это на рис. 33.

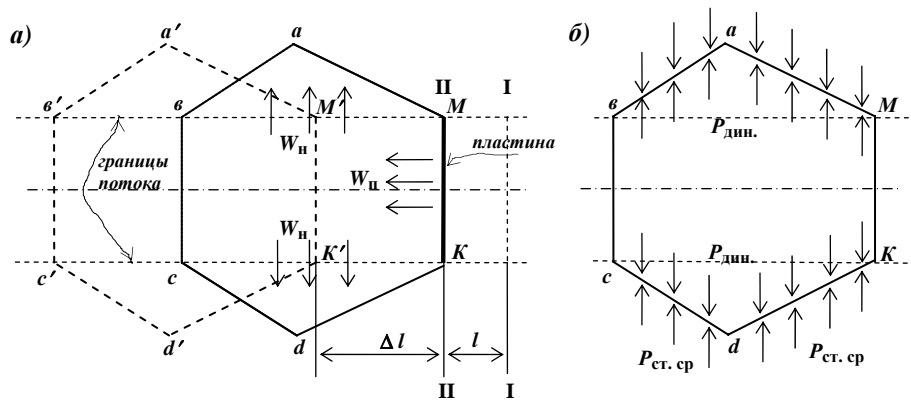

рис. 33

При своем перемещении пластина вытеснит определённый объём $V$, который поступает в окружающую среду. Считаем время останова пластины в положении II фиксированным.

К этому времени вытесненный объём газа займет в окружающей среде некий объём $a, в, с, d, K, M$, который мы назвали зоной возмущения. Он представляет собой объём тела вращения, осью которого служит ось потока. Поверхности объёма: $a, в, M$ и $с, d, K$ могут быть любой другой формы, чем той, которая показана на рис. 33. Форма определяется скоростью движения пластины возмущения. Здесь они показаны треугольной формы лишь для удобства изображения. Поступательные скорости движения и силы давления в этом объёме определяются зависимостями акустического вида движения.

Мы знаем, что образовавшийся объём $a, в, с, d, K, M$ с течением времени как-то должен изменяться. Чтобы определить это изменение, полагаем, что изменение происходит в течение небольшого промежутка времени $\Delta t$ от первоначального фиксированного времени. Через время $\Delta t$ берем второй фиксированный момент движения и смотрим, что произошло к этому времени.

К этому времени объём $a, в, с, d, K, M$ просто переместится на некоторое расстояние $\Delta l$ по прямой в направлении акустического движения и займет новое положение $a', в', с', d', K', M'$. При этом величина и форма останутся неизменными. Если мы рассмотрим еще целый ряд фиксированных моментов, начиная от первого, то мы заметим, что возмущенный объём просто перемещается со скоростью возмущения или скоростью звука. При этом до и после движущегося объёма остается зона невозмущенной среды. Теперь мы можем объёмно представить себе акустический поток жидкости или газа.

Дальше мы должны уяснить, как образуется и каким способом сохраняется движущийся объём возмущенной жидкости или газа. Поступательный расход массы в единицу времени, который определяется движением пластины возмущения, действует в объёме потока. Границы потока показаны на рис. 33, $a$. Этот расход вытесняет из объёма потока нормальный расход массы, который движется перпендикулярно границам потока. Поступательный и нормальный расходы массы в единицу времени равны между собой. Определяющим среди них является поступательный расход. Нормальные скорости зависят не только от поступательного расхода, но и от площади сечения. Для нормального потока она является переменной. За время движения пластины из положения I в положение II высота площади сечения потока увеличивается пропорционально скорости возмущения или скорости движения звука. И поступательное, и нормальное движение определяется зависимостями акустического вида движения. Мы просто вспомнили связь между поступательным и нормальным движением акустического вида движения.

Под действием нормальных динамических сил давления происходит изменение границ потока, например, до нового положения: *а, в, М* (рис. 33, *а*), так как они действуют нормально к границам потока. Новое положение границ *а, в, М* определяется тем, что в начальный момент на границе потока нормальные динамические силы давления бывают больше, чем статические силы окружающей среды. Поэтому границы зоны возмущения начинают увеличиваться. Но с ростом площади границ происходит уменьшение нормальных динамических сил давления. Это уменьшение находится в прямой зависимости от увеличения площади. Затем наступает такой момент, когда нормальные динамические силы давления станут равными статическим силам давления окружающей среды. В нашем примере такой момент наступает на границах объёма *а, в, М* и *с, d, K*. Таким образом определяются границы объёма возмущения в нормальном движению потока направлении, а изменение его границ в направлении движения нам уже известно. Зависимостей для определения нормальных границ возмущения в данной работе [т. е. «Механика жидкости и газа или механика безынертной массы. – *ред.*] нет, но их легко можно установить из условия равновесия статических сил давления и нормальных динамических сил давления. Следовательно, величина возмущённого объёма поддерживается динамическими силами давления, которые уравновешиваются статическими силами давления окружающей среды.

Динамические силы давления сохраняются в объёме за счет перемещения всего возмущенного объёма со скоростью возмущения. Так как для их существования необходим соответствующий расход массы в единицу времени. В конечном итоге состояние возмущённого объёма определяется избытком энергии по сравнению с окружающей средой, который получается за счет работы пластины возмущения.

Величина избытка энергии определяется уравнением работ V.46 [$L_\text{п} = v\rho \int_{W_{\text{п}1}}^{0} W_\text{п} dW$]. По этому уравнению мы можем получить ее количественную величину. Дальше для реальных жидкостей и газов мы обязаны записать избыток энергии уравнением термодинамической энергии сжатия VII.7 [$l = \int_{1}^{V_2} PdV$]. Так как он приводит к изменению термодинамического состояния среды. По уравнению работ V.46 мы вычислим величину энергии сжатия. Тогда по уравнению VII.7 останется определить пределы изменения объёма газа в возмущенной среде, то есть пределы интеграла этого уравнения. Для этого нам еще потребуется зависимость адиабатического процесса, так как избыток энергии в окружающей среде мы получаем за счет подвода в нее определённого количества массы в виде поступательного расхода массы в единицу времени. В этом примере мы подчеркнули другую особенность содержания избытка энергии в среде, которая отличается от избыточного содержания энергии набегающего потока.

Отметим, что для образования волны достаточно одного этапа движения пластины, например, из положения I в положение II. Хотя мы выше определили длину волны как образованную за два этапа движения пластины возмущения. Это было сделано для того, чтобы подчеркнуть различие в длине волн первого и второго этапов движения, а не в угоду ныне существующему понятию о длине волны, которое, пожалуй, было принято лишь в угоду синусу, а не самой природе акустического движения.

Если бы мы теперь начали рассматривать движение пластины из положения II в положение I, то нам бы пришлось повторить все вышеизложенные рассуждения с той лишь разницей, что они должны были бы относиться к возмущенному объёму с недостатком энергии относительно окружающей среды. В связи с этим скорости изменят свой знак на противоположный относительно скоростей возмущенного объёма первого этапа движения.

Мы несколько отвлеклись от задачи нашего примера. Мы просто должны найти нужную нам конструктивную схему глушителя и принцип его работы, так как движение газа при выхлопе соответствует акустическому движению, которое определяется как движение пластины конечной площади из положения I в положение II.

Дополнительная величина энергии возмущенного объёма относительно энергии окружающей среды создается путем ввода в этот объём дополнительной массы. Если убрать этот излишек массы, полностью или частично, то дополнительная энергия возмущенного объёма тоже либо исчезнет, либо останется частично.

Возьмем это положение в качестве принципа работы глушителя выхлопа газов. Следовательно, конструктивная схема глушителя должна быть такой, чтобы она могла обеспечить необходимый расход массы из возмущенного объёма. В конечном итоге необходимо убрать дополнительную энергию возмущенного объёма. Величина механической энергии равна произведению объёма на давление. Тогда количество этой избытка энергии мы можем определить как произведение объёма глушителя $V_\text{гл}$ на величину энергии единицы объёма возмущенной части среды. Величина энергии единицы объёма возмущенной зоны определяется уравнением V.47 [$\sum U_\text{п} = U_+ v\rho \int_{W_{\text{п}1}}^{0} W_\text{п} dW$]. Тогда изъятое количество энергии $U_\text{из}$ можно записать в такой форме:

$$U_{из} = V_{гл} \; (U + v\rho \int\limits_{W_{н1}}^{0} W_{н} \, dW \;) \qquad (44)$$

Величина изъятой энергии должна быть равна или быть меньше дополнительной величины энергии $U_{доп}$ возмущённого объёма. Количество дополнительной энергии $U_{доп}$ определяется как произведение объёма газа $V_{газ}$ на работу единицы объёма. Эта работа определяется уравнением V-46 [$L_{н} = v\rho \int\limits_{W_{н1}}^{0} W_{н} \, dW$ ].

Тогда количество дополнительной энергии $U_{доп}$ можно записать в таком виде:

$$U_{доп} = V_{газ} \, . v\rho \int\limits_{W_{н1}}^{0} W_{н} \, dW \; . \qquad (45)$$

Теперь, приняв энергию $U_{из}$, изъятую объёмом глушителя, равной или меньшей дополнительной энергии $U_{доп}$ , мы сможем по уравнению (44) определить объём глушителя $V_{гл}$, необходимый для изъятия определённого количества дополнительной энергии.

Мы получили нужный объём глушителя. Теперь мы должны оформить его конструктивно таким образом, чтобы он соответствовал принципу работы глушителя. Покажем на рис. 34 одну из схем конструктивного оформления глушителя. Непосредственно выхлопным патрубком глушитель соединён с выхлопным окном цилиндра поршневого двигателя. Объём глушителя образует прямоугольник, вытянутый вдоль оси. В стенках глушителя просверлено множество отверстий, которые прикрываются лепестковыми клапанами. Задача этих лепестковых клапанов заключается в том, чтобы выпускать газ из объёма глушителя и не пропускать газ в его объём. На рис. 34 показано, что лепестковый клапан имеет ограничитель хода. Это несущественное дополнение. Рассмотрим работу глушителя.

При выхлопе газы попадают в объём глушителя через выхлопной патрубок. Через площадь сечения выхлопного патрубка плоскости $S_1$ выхлопные газы будут поступать в объём глушителя как поступательный расход массы в единицу времени для акустического вида движения при выталкивании некоторого объёма пластиной возмущения. Этот расход начнёт выталкивать в нормальном направлении соответствующий расход массы газа, который к этому моменту заполнял объём глушителя.

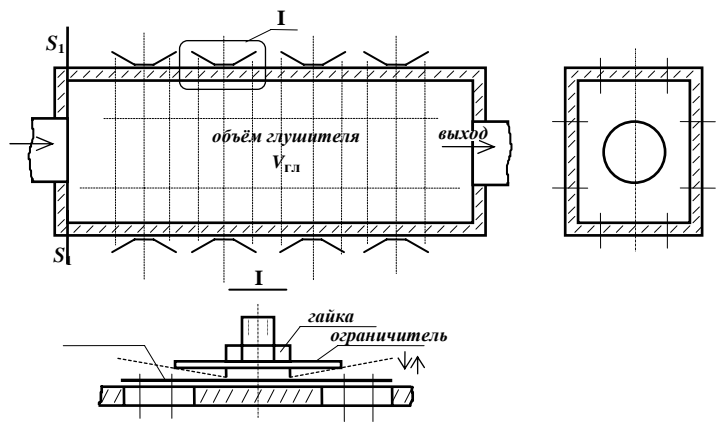

Рис. 34

Через отверстия в стенках глушителя нормальный расход начнёт поступать в окружающую среду. Клапана не препятствуют этому расходу. В результате образуется возмущённый объём среды, частью которого является объём глушителя. Возмущённый объём начнёт перемещаться в среде в соответствии с направлением акустического движения. Часть возмущённого объёма, которая заключена в объёме глушителя, останется без движения, так как лепестковые клапана закрываются и прерывают связь этих объёмов. Поэтому избыточная часть энергии остаётся в объёме глушителя. Без дополнительной энергии возмущённый объём перестаёт быть возмущённым объёмом. Следовательно, при выхлопе в зоне глушителя образуется акустический объём возмущённой среды, но он не перемещается в пространстве, Так как глушитель лишает его дополнительной энергии. Как видим, конструктивная схема глушителя (рис. 34) выполняет функции своей работы.

Остаётся определить остальные характеристики глушителя. Стенки глушителя должны располагаться как можно ближе к площади сечения входного патрубка или выхлопного окна цилиндра. После этого определяется длина глушителя.

Общая площадь отверстий в стенках глушителя определяется расходом массы выхлопных газов. Этот расход попадает в объём глушителя и выталкивает нормальный расход массы, который по величине равен расходу массы выхлопных газов. По уравнениям акустического вида движения вычисляются нормальные скорости и нормальные динамические силы давления. По величине нормального расхода и нормальной скорости определяются динамические силы давления. По величине нормального расхода и нормальной скорости с помощью уравнений движения определяется площадь всех отверстий в стенках глушителя.

Нормальные динамические силы давления открывают лепестковые клапана. Поэтому силу упругости лепестковых клапанов надо рассчитывать на их действие. Проделав эти несложные расчёты, мы определим необходимые характеристики глушителя.

Отметим, что работа глушителя подобного типа не уменьшает мощность поршневых двигателей, как это делают глушители современных конструкций.

В конце 70-х годов автор собрался сделать гребной винт новой конструкции для моторной лодки с мотором «Вихрь». Чтобы отработать технологию расчёта движителя «ТОЛИК» и его изготовления, автор сделал варианты расчётов и при помощи обыкновенного сверла изготовил несколько моделей.

Изготовить движитель в заводских условиях не удалось. Его обрабатывали на фрезерном станке. Он уже был готов процентов на 60%, когда редактор, работавший тогда на заводе, в очередной раз выносил полуобработанную заготовку с завода для дальнейшей поэтапной разметки и консультаций. Его остановила охрана, и больше он не видел заготовку. Хотя «дело» разбирал «товарищеский суд», но товарищи не вняли просьбам позволить оплатить металл, работу фрезеровщика и рассуждениям о пользе для науки и техники. Впрочем, возможно, потеря не велика. Автор говорил, что посадочное место у двигателя «Вихрь» слишком мало для рассчитанного им движителя. Так как лодка была чужая, то изменить посадочное место было нельзя. Поэтому автор пересчитывал, укорачивал лопасти, увеличивал их количество или ширину, в ущерб совершенству, т.к. нельзя было полностью сохранить нужные характеристики, и, видимо, остановился на каком-то компромиссном варианте, который, как сказано, не был осуществлён.

Ниже приводится фотография моделей, сделанных автором в домашних условиях.

Редактор

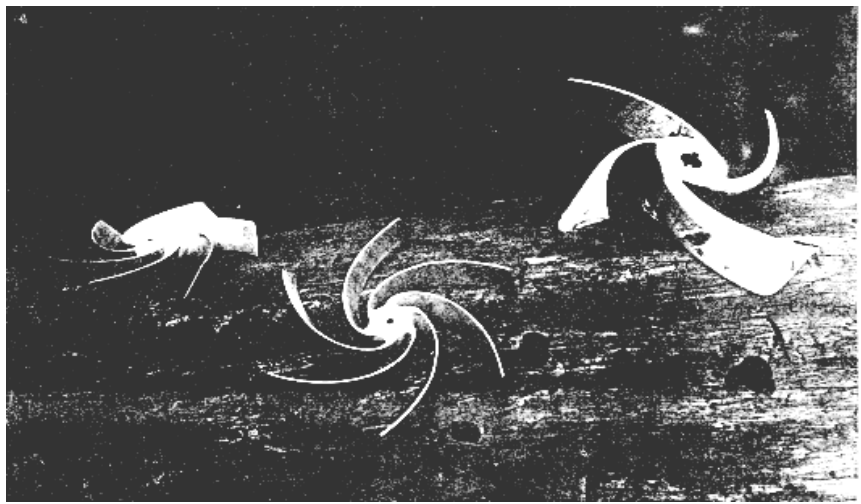

*От редактора*

Недавно редактор прочитал книгу Галилео Галилея «Пробирных дел мастер», откуда видно, что история аккуратно повторяется. Галилео Галилей написал эту книгу, чтобы дать отпор некоему рецензенту и «толкователю» его слов по фамилии Сарси. Так вот, этот Сарси тоже, между прочим, не видел различия между твёрдым и жидким «телом», а Галилей видел. Видел факт безынертности. Книга написана в 1623 году. Если бы сейчас Галилей и Сарси воскресли, то первый был бы за получить объяснения своим наблюдениям в виде механики безынертной массы, а второй, как в подобном не нуждался тогда, не стараясь понять труды Галилея, так и теперь не стал бы ничего понимать. Но только автор теории механики безынертной массы, но и сам великий Галилей потерял драгоценное время на опровержения измышлений «оппонентов», подобных Сарси, чтобы просто хотя бы сохранить свои мысли от искажения, без надежды просветить. Спрашивается, какая тогда польза науке и обществу от людей подобных Сарси? Слышал ли кто об этом сеньоре и его достижениях? Галилей ненароком увековечил имя Лотарио Сарси в своей книге. Впрочем, речь не о Сарси, а о безынертности, которую в упор не замечают, оказывается, в течение почти 400 (!) лет. Даже такие люди, как Д. Бернулли, всё-таки считали особенности движения жидкостей и газов видоизменённым движением твёрдого тела, а не самостоятельным видом движения массы. Но они, в отличие от критиков, сделали неоценимый вклад в науку. Разница в том, что теперь, когда свойство безынертности сформулировано, должно, наконец, появиться осознанное понимание различия движения. Если бы оно было у Д. Бернулли, то его вклад был бы особенно больше, так же как и у других таких, как он, людей. Из этого отрывка будет также видно, что открыватель принципа относительности движения – Галилей, не доводил свой принцип до абсурда, не теряя чувства реальности, как, якобы, его последователи.

«<...> К сказанному надлежит добавить ещё одно измышление, против которого Сарси, по его словам, не имеет возражений и даже называет его тождеством: по его мнению, утверждать, что жидкое тело, заключённое в полости сферического твёрдого тела, может быть увлечено последним, если твёрдое тело приведено во вращение, означает утверждать столь же легко и естественно, что и твёрдое тело, заключенное в жидкую субстанцию, было бы вовлечено в движение, если бы жидкость вращалась.

Утверждать такое – все равно, что думать, будто подобно тому, как судно переносится и увлекается течением реки, так и вода в стоячем пруду должна увлекаться вслед за движением лодки, что совершенно неверно. Прежде всего, сошлёмся на опыт: мы знаем, что течение реки приводит в движение судно или даже тысячу судов, заполняющих всю реку, в то время как судно, с какой бы скоростью оно ни плыло, не увлекает за собой воду. <...>

Так тяжёлый груз, подвешенный на верёвке, сохраняет в течение многих часов импет и движение, переданные ему одним толчком; вместе с тем можно сколь угодно сильно привести в движение воздух в комнате, но, как только импет[3] вызвавший движение, перестает действовать, воздух полностью успокаивается, и переданное ему движение бесследно исчезает.

Если окружающее и движущееся тело – жидкость, и оно действует с силой на находящееся в нём твёрдое, массивное и тяжёлое тело[4], то такое тело передает свое движение объекту, который в силу своей природы особенно пригоден для получения и сохранения движения в течение долгого времени. Это означает, что второй импет, идущий вслед за первым, налагается на движение, уже переданное твёрдому телу; третий импет налагается на движение, переданное телу первым и вторым импетом; четвертый импет добавляется к действию, произведенному первым, вторым и третьим импетом; так, шаг за шагом[5], – каждый последующий импет, сколько их ни будет, налагается на предыдущий, благодаря чему движение твёрдого тела не только не сохраняется, но и получает приращение. Но если движимое тело – жидкость, лёгкая и тонкая и, следовательно, не сохраняющая, а тотчас же утрачивающая передаваемое ей движение, то пытаться сообщить ей скорость – занятие столь же бесплодное, как пытаться наполнить бочку Данаид, которая мгновенно опорожняется, стоит лишь её наполнить водой. Смотри же, синьор Лотарио, сколь велика разница между двумя случаями, которые ты считал тождественными <...> ».

Остается добавить, что наблюдательность Галилея (в глазах редактора) – сверхъестественная, ибо он приводит для иллюстрации мгновенной, без отрицательного термина, потери движения легенду о бочке Данаид, которая опорожнялась мгновенно. Не постепенно вытекала, и даже не чрезвычайно быстро, давая таким образом хоть краткий отдых наполнявшим её, а именно мгновенно, т.е. с какой-то величины - до нуля, без промежуточных значений уровней. Поэтому, если бы определённое динамическое давление в трубопроводе могло бы сразу превратиться в 0, без промежуточных изменений своего значения, то и скорость жидкости сразу бы стала нулевой, без гидроударов (любая запорная арматура уменьшает площадь, т.е. меняет давление постепенно), без торможения. Ибо его для жидкости не существует (оно – кажущееся, похожее), так как её скорость соответствует действующим силам. Не зная этого свойства, нельзя, например, понять акустический вид движения, и остальные виды тоже.

---

[3] impetus (лат.) – побудительная сила движения. – *пояснение, взятое из книги*. Впрочем, перевод этой книги надо будет уточнять, чтобы не было бессмысленной тавтологии типа: побудительной силой движения является сила. В общем, Галилей имеет в виду причину силы, возможно, энергию, ибо сама сила наглядно есть действие.

[4] Галилей уже различал массивность и тяжесть!

[5] Обратите внимание, что Галилей утверждает дискретность существования действия сил во времени и дискретность существования причины силы. И дискретностью объясняет ускорение.